\documentstyle[preprint,epsfig,aps]{revtex}

\newcommand{\ba}{\mbox{$\bf a$}}
\newcommand{\bb}{\mbox{$\bf b$}}
\newcommand{\bd}{\mbox{$\bf d$}}
\newcommand{\br}{\mbox{$\bf r$}}
\newcommand{\bR}{\mbox{$\bf R$}}
\newcommand{\bt}{\mbox{$\bf t$}}

\newcommand{\bG}{\mbox{$\bf G$}}
\newcommand{\bk}{\mbox{$\bf k$}}

\newcommand{\calH}{\mbox{$\cal H$}}

\newcommand{\beq}{\begin{equation}}
\newcommand{\enq}{\end{equation}}

\newcommand{\hbttm}{\mbox{$\frac{\hbar^2}{2m}$}}
\draft

\begin{document}
\title{Quantum Confinement Effects in Semiconductor Clusters II}
\author{Antonietta Tomasulo and Mushti V. Ramakrishna}
\address{The Department of Chemistry, New York University,
New York, NY 10003-6621}
\date{Submitted to J. Chem. Phys. \today}
\maketitle

\begin{abstract}

The band gaps and spectral shifts of CdS, CdSe, CdTe, AlP, GaP, GaAs,
and InP semiconductor clusters are calculated from band structure
calculations using accurate local and non-local empirical
pseudopotentials.  The effect of spin-orbit coupling on the band
structures is included in the calculations when they are important.
The complete set of pseudopotential parameters and full computational
details are reported for all these semiconductors.  The calculated
spectral shifts of zinc-blende  and wurtzite CdS, wurtzite CdSe, and
zinc-blende InP  clusters are in good agreement with experiments
over a range of cluster sizes.  The effect of crystal structure on the
band gaps is small in large clusters but becomes important in small
clusters.  In the absence of experimental data, our calculations
provide reasonable estimates of expected spectral shifts for the other
clusters.  These results demonstrate that the empirical pseudopotential
method yields unique insights into the quantum confinement effects and
is a powerful tool for calculating the spectral shifts of semiconductor
clusters.

\end{abstract}

\section{Introduction}

Clusters are an embryonic form of matter whose microscopic study
provides insights into the evolution of material properties from
molecules and surfaces to solids
\cite{Pool:90,Corcoran:90,Bjornholm:90}.
Furthermore, clusters have been
shown to exhibit exotic optical properties and reactivities quite
different from those in molecules and solids
\cite{Brus:86,Wang:87,Elkind:87}.  For these reasons, theoretical
studies on clusters are critical to the design and synthesis of
advanced materials with desired optical, electronic, and chemical
properties
\cite{Kax:88,Kax:89,Bolding:90,Jelski:91,Proot:92,Delley:93,ONeil:90,Nair:92,Zhang:93,Hill:94}.
Such studies are at the interface of the traditional fields of quantum
chemistry, solid state chemistry, and statistical mechanics
\cite{Bawendi:89,Shiang:90,Hu:90,Canham:90}.  Hence, physicists,
chemists, and material scientists are working individually and in teams
to unearth the fundamental principles underlying the structure,
dynamics, and reactivities of these clusters
\cite{Krishnan:85,Biswas:85,Tersoff:86,Tomanek:87,Bawendi:90,Menon:91}.
However, theoretical studies on the spectroscopy of semiconductor clusters have
lagged far behind \cite{Yoffe:93}.

The interest on the spectroscopy of semiconductor clusters arose from
the discovery by Louis Brus that such clusters can be synthesized in
colloidal suspensions by controlled liquid phase precipitation
reactions \cite{Rossetti:83}. The radii of the clusters thus synthesized
are usually in the 5-100 \AA~ range.  Furthermore, the X-ray and
transmission electron microscopy (TEM) experiments have shown that
these clusters have the same lattice structures as the corresponding
bulk materials \cite{Brus:86,Wang:87,Bawendi:89,Shiang:90,Brus:93}.
Finally,
it is possible to prepare macroscopic samples of these clusters either
in powder form or in colloidal suspension form.  Such a versatility in
sample preparation has made it possible for experimentalists to
investigate the physical and optical properties of these clusters  in
detail \cite{Brus:86}.  Indeed, very sophisticated experiments are
currently underway in various laboratories to understand the absorption
spectra of these clusters as a function of the cluster size
\cite{Wang:87,Bawendi:89,Shiang:90,Hu:90}.  What is remarkable about
these clusters is that the electronic spectra of these clusters are not
the same as that of the bulk, even though the clusters have the same
structural properties as the crystals \cite{Brus:86}.  The
absorption spectra of clusters show relatively sharp resonances
superimposed on a continuum.  The first absorption peak in the spectrum
corresponds to the threshold  for the absorption of light by the
semiconductor cluster. It corresponds to the exciton energy, the
energy needed to excite an electron from the top of the valence to the
bottom of the conduction band.  Quantitative prediction of the shift of
the exciton energy with cluster size had been an outstanding problem
for a decade.  Simple theoretical calculations based on an effective
mass model (EMM) were not successful and {\em ab initio} quantum
chemistry electronic structure calculations are impossible for these
clusters consisting of thousands of electrons.  Consequently, what was
needed was a simple model that captured the essential physics of the
problem at hand.

Recently, we developed such a model and solved this problem
\cite{RK:91}. We used accurate pseudopotentials to carry out band
structure calculations and obtained the electronic energy levels of
these clusters.
Our calculations of the threshold for the absorption of light by CdS
clusters yielded results in excellent agreement with experiment over a
range of cluster sizes \cite{RK:91}.  Furthermore, we predicted two new
effects.  First, we found that the exciton energies in small CdS
clusters are sensitive to the crystal structure, even though such
a sensitivity is absent in large clusters \cite{RK:91}.  Second, we
found that the vertical Franck-Condon transition energies in indirect
gap clusters exhibit an anomalous redshift in small  clusters
\cite{RK:91}.  In sharp contrast, the extant theoretical models always
yielded monotonic blueshift of the transition energies with decreasing
cluster size \cite{RK:91}.

The band structure model we employ here has also yielded a
noteworthy insight into the nature of the electronic transitions in
semiconductor clusters.  The question is whether the electronic
transitions observed in these clusters are related in any way
to the bulk transitions.  Previous theoretical calculations
implicitly assumed that the electronic states in semiconductor clusters
are similar to those in molecules and hence they are qualitatively
different from those found in bulk.  In sharp contrast, our
band structure model underlies the presumption that the electronic
states in nanoscale clusters are similar to those in bulk.
Consequently, we use the bulk language (direct vs. indirect) to
classify the electronic transitions in clusters.  Recent experiments of
Brus and co-workers \cite{Brus:93} on the luminescence of Si clusters
support our assumption.  Based on the analysis of the luminescence of
Si clusters, Brus and co-workers concluded that the lowest energy
exciton transition is essentially an indirect gap transition,
eventhough the spectrum has blueshifted by almost 0.9 eV
\cite{Brus:93}.  This observation that bulk-like transitions are
preserved even in clusters supports our band structure model.

In this paper we extend our band structure calculations to a variety of
semiconductor clusters for two reasons:  First, we wish to investigate
the reliability of our band structure model by repeating these
calculations on different systems for which accurate experimental data
are now available.  Second, the accuracy of our calculations on
wurtzite clusters could not be verified previously because of the
absence of experimental data on those clusters at that time.  Since
highly accurate experimental data are now available for wurtzite CdS
and CdSe clusters, we felt the need to repeat our band structure
calculations on these clusters.  In this paper we also give the results
of our calculations on CdTe, AlP, GaP, GaAs, and InP clusters.  While
accurate and reliable experimental data on these clusters are not
available at present, we hope that our calculations provide reasonable
estimates of the expected spectral shifts as a function of cluster
size.

Experimental synthesis of semiconductor clusters is still a challenging
problem at present.  One major difficulty is establishing a suitable
synthetic route that yields nanometer scale clusters.  The other major
difficulty is determining optimal control parameters that yield
clusters with a narrow size distribution.  While both these problems
remain to be fully solved, significant progress is being made daily.  We
hope the present paper will serve as a useful guide to the
experimentalists on the expected behavior of spectral shifts in a
variety of technologically important nanoscale semiconductor clusters.

This paper is organized as follows:  Section II  gives the empirical
pseudopotential theory of band structure calculations, Sec. III
presents the band structure model, Sec. IV gives some computational
details, and Sec. V presents results of these calculations on a variety
of semiconductor clusters.  Finally, we summarize this paper in Sec.
VI.

\section{Empirical pseudopotential method}

The electronic structure calculations of the semiconductor clusters
are carried out using the empirical
pseudopotential method (EPM) that has been previously used for the
investigations of the optical properties of bulk semiconductor
materials and clusters \cite{RK:91}.
This method consists of solving the Schr\"{o}dinger
equation using an empirically determined pseudopotential for the valence
electron,
\beq
\calH = -\hbttm\nabla^2 + V_p,
\enq
\beq
V_p(\br, E) = V_L(\br) + \sum_{l = 0}^{\infty} \Pi_l^{\dagger} A_l(E)
f_l(\br)\Pi_l,
\enq
where the first term $V_L$ is the purely local part, the second term gives
the non-local $(V_{NL})$ part, and $\Pi_l$ is the projection operator
for angular momentum states $l$.
The local part of the pseudopotential is given by
\beq
V_L(\br)
=  \sum_{\bG} \left[ V_S(\bG)S_S(\bG) + i V_A(\bG)S_A(\bG) \right]
\exp(i\bG\cdot\br),
\label{VL}
\enq
where $V_S$ and $V_A$ are the symmetric and anti-symmetric form
factors, respectively, determined by fitting them to the experimental
optical data.  Similarly, $S_S$ and $S_A$ are the symmetric and
anti-symmetric structure factors, respectively, determined from the
crystal structure.  The function
$A_l(E)$ is the energy dependent well depth
\beq
A_l(E) = \alpha_{l} + \beta_{l}{[E^0(K)E^0(K')]^{1/2} - E^0(K_F)]},
\enq
where $E^0(K) = \hbar^2 K^2/2m$ and $K_F = (6\pi^{2}z/\Omega_c)^{1/3}$
is the Fermi wave-vector with $z$ the number of
valence electrons per unit cell and
$\Omega_c$ the volume of the unit cell.
$f_l(\br)$ is conveniently taken to be the square well
\beq
f_l(\br) = \left\{ \begin{array}{ll}
                   1 \hspace{4.0 em} &  \br < R_l \\
                   0 \hspace{4.0 em} &  \br \geq R_l.
                   \end{array}
           \right.
\label{flr}
\enq
$\alpha_l, \beta_l,$ and  $R_l$ are the non-local parameters of the EPM theory
to be determined from the experimental optical data.

\subsection{Non-local Pseudopotentials}

In many instances the local part of the pseudopotential $(V_L)$ is
sufficiently accurate to represent the gross features of the band
structures correctly.  However, the non-local pseudopotential is much
more accurate over a broad range of energy scales.  In particular, the
calculated
band widths and band dispersion are closer to the experimental values
with the non-local calculations than with the local calculations.  The
non-local calculations are also much more satisfactory theoretically
since they incorporate the correct angular momentum dependence
nature of the pseudopotential
experienced by the valence electrons.

The local contribution to the pseudopotential is given
by Eq. (\ref{VL}).  Evaluation of these
matrix elements of the Hamiltonian will be described in Secs. IIC
and IID for zinc-blende and wurtzite structures, respectively.
In the plane wave basis the matrix elements of the
non-local pseudopotential are of the form
\begin{equation}
V_{NL}({\bf K, K'}) =
\frac{4\pi}{\Omega_a}\sum_{l,i}A_{l}^{i}(E)P_l(\cos(\Theta_
{KK'}))S^{i}({\bf K-K'})F_{l}^{i}(K,K'),
\end{equation}
as discussed elsewhere \cite{Chelikowsky:89}.

\subsection{Spin-Orbit Coupling in Band Theory}
In light elements the electron spin ($s$) and orbital angular momentum
($l$) are both good quantum numbers,  since the magnetic field
generated by the orbiting electron is too weak to induce coupling with
the electron spin.  However, in heavier elements the nearly relativistic
speed of the valence electron produces a sufficiently large
magnetic field that $l$ and $s$  are coupled, giving rise to
$j = l + s$ as the good quantum number.
The spin-orbit interaction thus couples electron dynamics
in spin and ordinary spaces, thereby reducing the overall
symmetry of the Hamiltonian.  This relativistic effect is represented
by the operator \cite{Bethe:57}
\begin{equation}
H_{SO} =  \frac{\hbar}{4m^2c^2} \sigma \cdot (\nabla V_p \times p),
\end{equation}
where $\sigma$ are the Pauli spin matrices, $V_p$ is the pseudopotential,
$p$ is the electron momentum, $m$ is the true electron mass,
and $c$ is the speed of light.
The matrix elements of the new Hamiltonian in the
plane wave representation are given by \cite{Walter:70}
\begin{eqnarray}
<{\bf k+G'},s'|{\bf H}|{\bf k+G},s> & = &
\hbttm |{\bf k+G}|^2\delta_{{\bf GG'}} \delta_{ss'} \nonumber \\
& + & S_S({\bf G-G'})[V_S(|{\bf G-G'}|^2)\delta_{ss'} -
i\lambda_S({\bf G'\times G}) \cdot \sigma_{ss'}] \nonumber \\
& + & i S_A({\bf G-G'})[V_A(|{\bf G-G'}|^2)\delta_{ss'} -
i\lambda_A({\bf G'\times G}) \cdot \sigma_{ss'}].
\end{eqnarray}
For a binary semiconductor consisting of two types of atoms
(A $\neq$ B), we define
\begin{equation}
\lambda_S = \frac{1}{2}(\lambda_1 + \lambda_2)$, $\lambda_A =
\frac{1}{2}(\lambda_1 -\lambda_2),
\end{equation}
where
\begin{equation}
\lambda_1 = \mu B_{nl}^{\rm A}(G)B_{nl}^{\rm A}(G')$,  $\lambda_2 =
\alpha\mu B_{nl}^{\rm B}(G)B_{nl}^{\rm B}(G'),
\end{equation}
$\mu$ is the adjustable parameter chosen in order to obtain
the splitting $\Delta$
of the valence band at $\Gamma$ correctly,  and $\alpha$ is the
ratio of the contribution from atom A to the contribution from atom B
at $\Gamma$.  The $B_{nl}$ are defined as
\begin{equation}
B_{nl}(G) = C \int_{0}^{\infty} r^2 R_{nl}(r) j_l(Gr) dr,
\end{equation}
where $j_l$ are the spherical Bessel functions,
$C$  is determined by the condition
\begin{equation}
\lim_{G \rightarrow 0} \frac{B_{nl}(G)}{G} = 1,
\end{equation}
and $R_{nl}$ are the radial parts of the outermost electron
wave functions taken from the Herman-Skillman tables
\cite{Skillman:63}.  For simplicity we always used a value of $\alpha =
1.0$ and $C$ = 1.0.

The band structures of semiconductors have several common features:
At $\Gamma$ the HOMO is split by the spin-orbit coupling
\cite{Harrison:80}.
Specifically, counting the spin, for the zinc-blende structure
the HOMO band is sixfold degenerate. The spin-orbit interaction
splits this  band into an upper fourfold degenerate $\Gamma_8$ component, and
a lower twofold degenerate $\Gamma_7$ component. The conduction band $\Gamma_1$
is twofold degenerate \cite{Phillips:73}.
For the wurtzite structure, the degeneracy of
the $\Gamma$ band is already removed by the crystal field,
leading to an upper twofold
degenerate  band, and a lower fourfold degenerate band. The spin-orbit coupling
then splits the lower band into two twofold degenerate bands
\cite{Phillips:73}.
For both systems, wurtzite and zinc-blende, the  highest three valence bands
are called heavy-mass, light-mass, and split-off
band, respectively \cite{Phillips:73}.
Transitions between
these three valence bands to the conduction band are traditionally labelled
A, B, and C \cite{Koch:93}.
The atomic orbital approach is especially useful for describing valence bands
near $\Gamma$. From this perspective,
the HOMO bands arise from the valence $p$ orbitals while the LUMO band
arises from the $s$ orbitals \cite{Phillips:73}.
In polar semiconductors such as CdS and
CdSe, HOMO bands originate primarily from the anion, and LUMO bands from
the cation.
However, in non-polar semiconductors such as CdTe and GaAs,
considerable mixing of the cation and anion orbitals will take place.
Since our interest is in reproducing the splitting of the valence $p$
bands near HOMO, our calculations considered only the
contributions from the outermost $p$ orbitals.  We do not consider
spin-orbit effects on the innercore or the $d$-electron states since
these states are eliminated in EPM through the use of the
pseudopotential.  The $s$ orbitals do not exhibit spin-orbit splitting
since their orbital angular momentum is zero.

\subsection{Local Pseudopotentials of Zinc-blende Structure Crystals}
The local pseudopotential $(V_L)$ experienced by a valence electron at a
point \br~ inside a crystal lattice is given by
\begin{equation}
V_L(\br) = \sum_{\bR, j} v_j(\br - \bR - \bd_j),
\end{equation}
where the summation is over all
the basis atoms $j$ at each lattice point \bR,
and $v_j$ is the atomic pseudopotential due to the $j$th basis atom at
the lattice site \bR.  Fourier expansion of $v_j$ yields,
\begin{equation}
V_L(\br) = \frac{1}{Nn_a} \sum_{\bG}\sum_{\bR, j} v_j(\bG)
\exp \left[ i\bG\cdot(\br - \bR - \bd_j) \right],
\end{equation}
where $N$ are the number of unit cells and $n_a$ are the
number of basis atoms per unit cell.

The zinc-blende (or sphalarite) lattice consists of two interpenetrating
fcc lattices, displaced from each other along the body diagonal by $a_0/4$,
$a_0$ being the lattice constant of the unit cell.
Consequently, we may view the zinc-blende structure
as an fcc lattice with two different A $\neq$ B
basis atoms per unit cell.  For this case,
$n_a = 2, v_1(\bG) \neq v_2 (\bG),$
and $\bd_1 = - \bd_2 = -\bt_1 = -(1,1,1)/8$.   Explicitly summing
the above equation now yields,
\begin{eqnarray}
V_L(\br) & = & \sum_{\bG} \left[ \frac{1}{n_a} \sum_j v_j(\bG)
\exp(-i \bG\cdot \bd_j) \right] \exp(i\bG\cdot \br) \\
&  = & \sum_{\bG} \left\{\frac{1}{2} \left[v_1(\bG) \exp(i\bG\cdot\bt_1)
+ v_2(\bG) \exp(-i\bG\cdot\bt_1) \right]\right\} \exp(i\bG\cdot\br) \\
& = & \sum_{\bG} \left[ V_S(\bG)S_S(\bG) + i V_A(\bG)S_A(\bG) \right]
\exp(i\bG\cdot\br), \label{VL2}
\end{eqnarray}
where
\begin{equation}
V_S(\bG) = \frac{1}{2}\left[v_1(\bG) + v_2(\bG)\right],~~~
V_A(\bG) = \frac{1}{2}\left[v_1(\bG) - v_2(\bG)\right]
\label{VsVa}
\end{equation}
are the symmetric $(V_S)$ and anti-symmetric $(V_A)$ form factors
and
\begin{equation}
S_S(\bG)  = \cos(\bG\cdot\bt_1),~~~
S_A(\bG)  = \sin(\bG\cdot\bt_1)
\end{equation}
are the symmetric $(S_S)$ and anti-symmetric $(S_A)$ structure
factors, respectively.
Defining the reciprocal lattice vectors \bG~ as
\begin{equation}
\bG = \frac{2\pi}{a_0}(G_x, G_y, G_z)
\end{equation}
we obtain
\begin{equation}
S_S(\bG) = \cos\left[\frac{\pi}{4}(G_x + G_y + G_z)\right],~~~
S_A(\bG) = \sin\left[\frac{\pi}{4}(G_x + G_y + G_z)\right].
\end{equation}

If we now specialize to the case of A = B, we obtain
the local pseudopotential for diamond structure:
\begin{equation}
V_L(\br) = \sum_{\bG} V_S(\bG)S_S(\bG) \exp(i\bG\cdot\br).
\end{equation}

\subsection{Local Pseudopotentials of Wurtzite Structure Crystals}

The wurtzite crystals are made of two interpenetrating hexagonal close
packed (hcp) lattices.   One hcp lattice is entirely made of A type
atoms and the other entirely of B type atoms (A $\neq$ B) and these two
lattices are displaced from each other by $ 2t_2 = \frac{3}{8} c_0$ along the
c-axis.  However, the hcp lattice is not a Bravais lattice.  The hcp
lattice consists of two interpenetrating simple hexagonal Bravais
lattices, displaced from one another by $ 2t_1 = (\frac{1}{3} \ba_1,
\frac{1}{3} \ba_2, \frac{1}{2} \ba_3)$, where $(\ba_1, \ba_2, \ba_3)$
are the direct lattice primitive translation vectors of the
simple hexagonal Bravais lattice.  Hence, the wurtzite structure
is a network of four simple hexagonal lattices, with four atoms per unit
cell and two different types of atoms.
For this case $n_a = 4$ and the position vectors of the four basis atoms
in the unit cell are given by
\begin{eqnarray}
\bd_1 & = & -(\bt_1 + \bt_2) \label{d1} \\
\bd_2 & = & -(\bt_1 - \bt_2) \label{d2} \\
\bd_3 & = &  (\bt_1 + \bt_2) \label{d3} \\
\bd_4 & = &  (\bt_1 - \bt_2). \label{d4}
\end{eqnarray}

The local pseudopotential of the valence electron interacting with a periodic
wurtzite lattice is given by
\begin{eqnarray}
V_L(\br) & = & \sum_{\bG} \left[\frac{1}{n_a}
\sum_{j}v_j(\bG)\exp(-i\bG\cdot\bd_j)\right]\exp(i\bG\cdot\br)
\label{Vpr4}\\
& = & \sum_{\bG}\frac{1}{n_a} \left\{v_1(\bG)\left[\exp(-i\bG\cdot\bd_1) +
\exp(-i\bG\cdot\bd_4)\right] \right.\nonumber\\
&&+ \left.v_2(\bG)\left[\exp(-i\bG\cdot\bd_2) + \exp(-i\bG\cdot\bd_3)
\right]\right\}\exp(i\bG\cdot\br),
\end{eqnarray}
since atoms 1 and 4,
and 2 and 3 are identical in the wurzite lattice.  Rewriting $v_1$
and $v_2$ in terms of $V_S$ and $V_A$ from Eq. (\ref{VsVa}) we obtain
\beq
V_L(\br) =
\sum_{\bG}\left[V_S(\bG)S_S(\bG) + i V_A(\bG)S_A(\bG)\right]\exp(i\bG\cdot\br),
\label{Vpr7}
\enq
where the symmetric $(S_S)$ and anti-symmetric $(S_A)$
structure factors are given by
\beq
S_S(\bG) = \frac{1}{n_a}\sum_j\exp(-i\bG\cdot\bd_j),~~~~~ S_A(\bG) =
\frac{-i}{n_a}\sum_j P_j\exp(-i\bG\cdot\bd_j), \label{SGhex}
\enq
where $P_j = +1$ when $j = 1, 4$ and $P_j = -1$ when $j = 2, 3$.  Thus
the pseudopotentials of wurtzite and zinc-blende crystals differ from
each other only in the definition of the structure factors.  We can
carry out the summations in Eq. (\ref{SGhex}) as follows.

The hexagonal lattice is characterized by three parameters:
$a_0$, $c_0$, and $u_0$.  Like zinc-blende, wurtzite
lattice has tetrahedral coordination about each ion, but the
orientation of the tetrahedron is different from that of zinc-blende.
If we assume perfect tetrahedral coordination, then
\beq
\frac{c_0}{a_0}  =  \sqrt{\frac{8}{3}},~~~~~ u_0  =  0.375 = \frac{3}{8}.
\enq
The direct lattice primitive translation vectors of the simple hexagonal
Bravais lattice are \cite{Ashcroft:76}
\beq
\ba_1  =  (1, 0, 0)a_0,~~~
\ba_2  =  (\frac{1}{2}, \frac{\sqrt{3}}{2}, 0)a_0,~~~
\ba_3  =  (0, 0, \frac{c_0}{a_0})a_0.
\label{a123}
\enq
With this definition, the position vectors of the basis atoms are
\begin{eqnarray}
\bd_1 & = & - \frac{1}{4}\left[1, \frac{1}{\sqrt{3}},
(1 + 2u_0)\frac{c_0}{a_0}\right] a_0  \label{bd1} \\
\bd_2 & = & - \frac{1}{4}\left[1, \frac{1}{\sqrt{3}},
(1 - 2u_0)\frac{c_0}{a_0}\right] a_0  \label{bd2} \\
\bd_3 & = &   \frac{1}{4}\left[1, \frac{1}{\sqrt{3}},
(1 + 2u_0)\frac{c_0}{a_0}\right] a_0  \label{bd3} \\
\bd_4 & = &   \frac{1}{4}\left[1, \frac{1}{\sqrt{3}},
(1 - 2u_0)\frac{c_0}{a_0}\right] a_0  \label{bd4}
\end{eqnarray}
in cartesian coordinate representation.

In the zinc-blende crystal the nearest neighbor atoms are located
half-way along the face diagonal of the fcc lattice.  Consequently, if
$a_0({\rm zb})$ is the lattice constant of the zinc-blende crystal, then the
nearest neighbor distance is
\beq
r_n ({\rm zb}) = \frac{1}{\sqrt{2}} a_0({\rm zb}).
\label{rnzb}
\enq
In the hexagonal crystal, the nearest neighbor atoms are located along
the edges of the hexagon.  Consequently,
\beq
r_n ({\rm hex})  =  a_0 ({\rm hex})
\label{rnhex}
\enq
Since $r_n ({\rm zb}) = r_n ({\rm hex})$, comparing Eqs.
(\ref{rnzb}) and (\ref{rnhex}) we obtain the relation
\beq
a_0 ({\rm zb}) = \sqrt{2} a_0 ({\rm hex}).
\label{a0zb}
\enq
and
\beq
\bG = \frac{\sqrt{2}\pi}{a_0({\rm hex})}(G_x, G_y, G_z).
\label{Geehex}
\enq
This definition allows the comparison
of \bG~ vectors of zinc-blende and wurtzite crystals on an equal footing.

Substituting Eqs. (\ref{d1}-\ref{d4}) into (\ref{SGhex}) and
carrying out some algebraic manipulations we obtain
\beq
S_S(\bG) =    \cos(\bG\cdot\bt_1) \cos(\bG\cdot\bt_2),~~~~~
S_A(\bG) =  - \cos(\bG\cdot\bt_1) \sin(\bG\cdot\bt_2),
\label{SSAhex1}
\enq
where
\beq
\bt_1 = \frac{1}{6} (\ba_1 + \ba_2) + \frac{\ba_3}{4}
=  (\frac{1}{4}, \frac{1}{\sqrt{48}}, \frac{1}{\sqrt{6}}) a_0,
\enq
\beq
\bt_2 = \frac{1}{2}u_0 \ba_3
=  (0, 0, \sqrt{u_0}/2) a_0.
\enq

Finally, substituting Eq. (\ref{Geehex}) into (\ref{SSAhex1}) we obtain
\beq
S_S(\bG) =   \cos\left[\sqrt{2}\pi\left(\frac{G_x}{4} + \frac{G_y}{\sqrt{48}}
 + \frac{G_z}{\sqrt{6}}\right)\right] \cos(\sqrt{2}\pi G_z \sqrt{u_0}/2),
\label{SSAhex2}
\enq
\beq
S_A(\bG) = - \cos\left[\sqrt{2}\pi\left(\frac{G_x}{4} +
\frac{G_y}{\sqrt{48}} + \frac{G_z}{\sqrt{6}}\right)\right]
\sin(\sqrt{2}\pi G_z \sqrt{u_0}/2).  \label{SSAhex3} \enq

We can also define \bG~ as
\beq
\bG = (l\bb_1, m\bb_2, n\bb_3),
\label{Geehexb}
\enq
where $\bb_1, \bb_2,$ and $\bb_3$ are the reciprocal lattice
primitive translational vectors given by
\beq
\bb_1  =  \frac{\sqrt{2}\pi}{a_0}(\sqrt{2}, -\sqrt{\frac{2}{3}}, 0),~~~~~
\bb_2  =  \frac{\sqrt{2}\pi}{a_0}(0, \sqrt{\frac{8}{3}}, 0),~~~~~
\bb_3  =  \frac{\sqrt{2}\pi}{a_0}(0, 0, \frac{\sqrt{2}a_0}{c_0}).
\label{b123}
\enq
Substituting Eqs. (\ref{b123}) and (\ref{Geehexb}) into (\ref{SGhex}),
utilizing the relations
\beq
\bb_i\cdot \ba_i = 2\pi,~~~ \bb_i\cdot \ba_j = 0~~~ (i\neq j),
\enq
and carrying out some algebraic manipulations, we obtain
\beq
S_S(\bG) = \cos\left[2\pi\left(\frac{l}{6} + \frac{m}{6} +
\frac{n}{4}\right)\right] \cos(n\pi u_0),~~~~~~
S_A(\bG) = \cos\left[2\pi\left(\frac{l}{6} + \frac{m}{6} +
\frac{n}{4}\right)\right] \sin(n\pi u_0).
\label{SSAhex4}
\enq
Eqs. (\ref{SSAhex2})--(\ref{SSAhex3}) and (\ref{SSAhex4}) are
equivalent.  We find that the above structure factors yield band
structures in excellent agreement with those of Cohen and Bergstresser
for wurtzite crystals \cite{Berg:67}.  We tested our program with both
these structure
factors and verified that they give identical band structures.

Previously we used a different coordinate system for the definition of
the direct lattice vectors of the simple hexagonal Bravais lattice.
That definition resulted in structure factors different from the ones
given above.  While these two definitions are equivalent in principle,
in practice the values of the
parameters $V_S$ and $V_A$ are intimately linked to the
choice of the coordinate system and the structure factors
determined therefrom.  The present choice of the coordinate system that
gives rise to the structure factors in Eqs.
(\ref{SSAhex2})--(\ref{SSAhex3}) and (\ref{SSAhex4}) yield correct
band structures, in complete agreement with the original band
structures of Cohen and Bergstresser \cite{Berg:67}.

\section{Band Structure Model}

The virtue of EPM is that it reproduces the bulk band structure to
better than 0.1 eV accuracy \cite{RK:91}.  Other methods are less
accurate, especially for the calculations of the band gaps.  For
example, the calculations based on the density functional theory (DFT)
typically underestimate the band gaps by about 30-50\%.  The DFT based
methods optimize the orbitals of the occupied electronic states only,
not those of the unoccupied orbitals.  This problem remains even when
non-local density gradient corrections and other higher-order
improvements to the DFT methodology are made.  Consequently, the band
gaps determined by DFT are in general grossly in error.  Furthermore,
the DFT calculations are computationally far more expensive compared to
EPM.  Hence, EPM is suitable for the investigation of the electronic
structures of semiconductor clusters.

To apply EPM to the electronic structure calculations of the
semiconductor clusters we assume that these clusters have the crystal
structure of the bulk semiconductor.  This assumption is justified
because we are considering relatively large clusters containing tens to
hundreds of atoms.  Furthermore, the X-ray and TEM experiments on Si,
CdS, CdSe, CdTe, GaAs, and InP clusters have shown that the bulk
lattice structure is preserved even when the cluster radii are as small
as \bR~ = 7 \AA
\cite{Brus:86,Wang:87,Bawendi:89,Shiang:90,Hu:90,Bawendi:90}.  The
reason for the preservation of the bulk lattice structure in such small
clusters may be due to the presence of ligands on the surfaces of these
clusters.  These ligands are necessary to prevent the clusters from
coalescing into larger units.  These ligands also terminate the
dangling bonds on the surfaces of these clusters, and thus inhibit
structural reconstruction.  For this reason, even small clusters seem
to possess bulk lattice structure.   The major effect of size on
cluster structures seems to be small contractions of the bonds relative
to their bulk bond lengths.

The band
structure calculations are carried out for these clusters in almost the
same way as we had done previously for CdS \cite{RK:91}.
In bulk semiconductors the allowed wave vectors \bk~ of the
electronic states are continuous.  On
the other hand, only discrete $\bk$-states are allowed in clusters.  If
we model the cluster as a rectangular box with dimensions $L_x, L_y$,~
and $L_z$, then a reasonable
approximation is to use
the bulk pseudopotential $V_p(\br)$ inside the
cluster and terminate this potential at the surfaces of the cluster by
an infinite potential.
The wave vectors of the lowest
allowed states are then given by the
quantization condition $\sin(k_xL_x)\sin(k_yL_y)\sin(k_zL_z) = 0$,
whose solution is
\beq \bk  =
\pi\left(\frac{n_x}{L_x},\frac{n_y}{L_y},\frac{n_z}{L_z}\right),
\label{krect}
\enq
where $n_x$, $n_y,$ and $n_z$ are the quantum numbers of a particle in
a box with infinite potentials at the boundaries.  The surface ligands
act a potential barriers to the valence electrons.
Consequently, for low energy excitations under consideration, the
assumption of infinite potentials at the boundaries is a good
approximation.  The energy levels at these allowed \bk-states
constitute the band structure of a rectangular box.

Similarly, if we model the cluster as a spherical object
of radius $R$, the energy levels of the valence electrons will be
quantized because of the spherical boundary.  The wave vectors of the
lowest allowed states are given by $j_0(k_nR) = 0$, whose solution is
$\bk_n = n\pi/R$ \cite{Flugge:71}.  Since $\bk_n$ is along the radial
direction, we project it onto each of the cartesian axes with equal
magnitude to obtain cartesian components of \bk.  This procedure yields
\beq
\bk  = \frac{\pi}{\sqrt{3}R}\left(n_x,n_y,n_z\right).
\label{ksphere}
\enq
for spherical clusters.  Sometimes we also model spherical clusters
as cubic boxes with $L = 2R$ as the sidelength of the cube.
The energy levels at these allowed \bk-states constitute the band structure
of a spherical cluster.

The exciton energy of a cluster of radius $R$ is given by \cite{RK:91}
\beq
E_x = E_g - \frac{1.786}{\epsilon R} - 0.248E_{Ry},
\enq
where $E_g$ is the band gap, $\epsilon$ is the dielectric constant,
$E_{Ry} = \mu e^4/2\epsilon^2\hbar^2$ is the effective Rydberg
energy of the exciton, and $\mu$ is the reduced mass of the electron-hole
pair.

EPM has been shown to reproduce the band gaps to within 0.1 eV for
both bulk materials and clusters  \cite{RK:91}.  Specifically,
our predicted indirect band gap of 0.43 eV for a 10 \AA~ radius silicon
cluster was found to be in good agreement with recent experimental
value of 0.5 eV obtained by Louis Brus and co-workers
\cite{Littau:93}.  Likewise, EPM yielded excellent results for CdS
clusters in comparison with experiment \cite{RK:91}.  Full details on
the computational methodology are given in Refs.  \cite{RK:91}.
The utility of the proposed method for the calculations of the spectral
shifts is also documented in the previous publications
\cite{RK:91}.

\section{Computational Details}

The binary semiconductors being considered in this study all consist of
either zinc-blende or wurtzite lattices.  Consequently, the Hamiltonian
matrices are of complex valued Hermitian matrices.  We diagonalized
these matrices using the EISPACK routines CH, HTRIDI, and TQLRAT.
Typically, we use 283 plane waves to converge the energies to better
than 0.01 eV accuracy.

Table I lists the parameters of the local pseudopotentials of
zinc-blende and wurtzite CdS and CdSe clusters, while Tables II and III list
the corresponding parameters for the non-local pseudopotentials of
zinc-blende CdTe, AlP, GaP, GaAs, and InP clusters.  Spin-orbit
interaction was included in the calculations of wurtzite CdS, wurtzite
CdSe, zinc-blende CdTe, and InP clusters.  We have verified that the
spin-orbit effects are small in AlP, GaP, and GaAs crystals.
Consequently, our calculations on these clusters omitted the spin-orbit
interaction in the Hamiltonian.

The $s$ orbitals don't exhibit spin-orbit interaction since they have
$l = 0$ orbital angular momentum.  The innercore and $d$ electron
states are eliminated in EPM through the use of the pseudopotentials.
Consequently, only the outermost $p$ orbitals are affected by the
spin-orbit interaction.
Hence, we used the radial Hartree-Fock-Slater orbitals
$4p$ for Cadmium, $5p$ for Indium, $3p$ for Sulphur, $4p$ for Selenium,
$5p$ for Tellurium, and $3p$ for Phosphorus in our calculations employing
spin-orbit interaction in the Hamiltonian.  Other calculations
reported in the literature seem to include innercore
$p$ orbital wave functions
in the Hamiltonian, instead of the outermost $p$ orbitals we employ.  However,
we have verified that in each
case we obtain
better band structures with the outermost $p$ orbital wave functions
than with the innercore $p$ orbitals.

We use the standard notation to represent the high symmetry points in
the Brillouin zone \cite{RK:91}.    For zinc-blende crystals
\beq
X = (1,0,0), W = (1,0.5,0), L = (0.5,0.5,0.5),
\Gamma = (0,0,0), U = K = (0,0.75,0.75)
\enq
in units of $2\pi/a_0$.  For wurtzite crystals, the high symmetry points
are defined as
\begin{eqnarray}
\Gamma & = & (0,0,0), K = (\sqrt{2}/3, \sqrt{2}/3, 0),
M = (\sqrt{2}/3, 0, 0), \nonumber \\
A & = & (0, 0, a_0)/\sqrt{2}c_0) = (0, 0, \sqrt{3}/4), \nonumber \\
H & = & (\sqrt{2/3}, \sqrt{2/3}, a_0/\sqrt{2}c_0)
  = (\sqrt{2/3}, \sqrt{2/3}, \sqrt{3}/4), \nonumber \\
L & = & (\sqrt{2/3}, 0, a_0/\sqrt{2}c_0) = (\sqrt{2/3}, 0, \sqrt{3}/4)
\end{eqnarray}
in units of $\sqrt{2}\pi/a_0$.  We map bands along these symmetry points
to obtain the complete band structure.

\section{Results and Discussion}

\subsection{CdS and CdSe Clusters}

The CdS and CdSe crystals exist in both zinc-blende and wurtzite form,
with the wurtzite form being the ground state structure.  In principle
it is possible to synthesize clusters also in both these lattice
forms.  However, the CdS clusters seem to prefer zinc-blende over
wurtzite, while CdSe clusters seem to prefer wurtzite over zinc-blende
structure.  The possibility of being able to synthesize both these
isomeric forms of CdS and CdSe clusters, provides an opportunity to
investigate the effect of lattice structure on the exciton energies as
a function of cluster size.

We use the lattice constant $a_0$ = 5.82 \AA~
for zinc-blende and $a_0$ = 4.14 \AA~ for wurzite CdS crystals.
Our calculated band gaps are 2.44 eV and 2.52 eV for
bulk zinc-blende and wurtzite CdS, respectively. The experimental
result is 2.50 eV \cite{Kayanuma:88} for both these structures.
We use the lattice constant $a_0$ = 6.08 \AA~
for zinc-blende and $a_0$ = 4.30 \AA~ for wurzite CdSe crystals.
The calculated band gaps for CdSe are 1.92 eV and 1.79 for bulk
zinc-blende and wurtzite CdSe, respectively.  The experimental values
for bulk wurtzite CdSe are 1.75 and 1.83 eV at 80 and 293 K,
respectively.  We take the average of these two values, 1.79 eV, as the
experimental value for both zinc-blende and wurtzite CdSe crystals.
According to Cohen and Bergstresser \cite{Berg:67}
the splitting in energy of the fundamental gap without spin-orbit
coupling $(\Gamma_6^v \rightarrow \Gamma_1^c)$ is $E_a + \frac{1}{3}
\Delta$, where $E_a$ is the value of the band gap with spin-orbit coupling
($\Gamma_9^v \rightarrow \Gamma_7^c$) and $\Delta$ is the value of
the spin-orbit
splitting. The spin-orbit coupling parameter $\mu$ was chosen in order to
satisfy this condition to better than 0.001 eV accuracy.
For CdS crystal, $\Delta =
0.062$ eV at 77K \cite{Semi:92} and we determined that $\mu$ = 0.00008
is optimal.
For CdSe
crystal, $\Delta$ = 0.42 eV at 77 K \cite{Semi:92} and we determined
that $\mu$ = 0.00041 is optimal.
Figure 1(a) show the band structure of a wurtzite CdS crystal, including the
spin-orbit coupling.
The discrete energy levels of a $R$ = 15.0 \AA~
zinc-blende CdS cluster,
modelled as a sphere of radius R, are given in Fig. 1(b).
Figure 1(c) show the band structure of a zinc-blende CdS zinc-blende crystal,
compared with experimental data \cite{Duke:88}.
Figure 2(a) show the band structure of a wurtzite CdSe crystal, including the
spin-orbit coupling, and 2(b) is the band structure of a zinc-blende CdSe
crystal,
compared with the available experimental data \cite{Duke:88}.
The study of the dependence of the excitonic energies on the cluster size
was carried out using the
bulk crystal parameters  $\epsilon$ = 5.5,
$m_e$ = 0.19,  $m_h$ = 0.80 \cite{RK:91} for both wurtzite and
zinc-blende CdS system, and
$\epsilon$ = 10.0, $m_e$ = 0.13,
$m_h$ = 0.45 \cite{RK:91} for both wurtzite and zinc-blende CdSe system.
The exciton energies are reported in Tables IV--VII.
Since for a cluster of radius $R = 100$ \AA~ the fundamental
gap is 2.46 eV  for zinc-blende, and
2.55 eV  for wurtzite, we shifted
our conduction energy levels by +0.04 eV and -0.05 eV, respectively,
in order to obtain the experimental bulk band gap of 2.50 eV.
We carried out a similar correction of
for CdSe clusters,  shifting the
conduction bands by -0.15 eV for zinc-blende, and -0.025 eV for
wurtzite, so that we obtain the bulk band gap of
1.79 eV for a cluster of $R = 100$ \AA.  In Fig. 3 we compare the
calculated exciton energies of both wurtzite and zinc-blende CdS
clusters with the experimental data
\cite{Rossetti:83,Rossetti:84,Chestnoy:85,Wang:90,Weller:94}.  This
figure clearly shows that the zinc-blende and wurtzite clusters exhibit
different spectral shifts and the available experimental data follow
one or the other of these trends.  Based on these calculations, we can
assign the zinc-blende structure to the clusters synthesized by Wang
and Herron \cite{Wang:90} and wurtzite structure to the smaller
clusters synthesized by Weller and coworkers \cite {Weller:94}.

\subsubsection {The Effect of Dielectric Constant and
Spin-Orbit Coupling}

We have investigated the dependence  of the
exciton energies on dielectric constant $(\epsilon)$ by carrying out
calculations on CdSe clusters using two different values of $\epsilon$:
$\epsilon(0)$ and  $\epsilon$($\infty$).  Figure 4(a) compares the
calculated exciton energies with the available experimental data
\cite{H-Borr:87,Peyg:89,Brus:90,Eki:93,Baw:93,Alivisatos:93}.
The dashed line represents the calculations on wurtzite CdSe clusters with
$\epsilon(0)$ = 10.0, while the solid line represents the corresponding
calculation with
$\epsilon$($\infty$) = 6.25.   According to these results the exciton energies
calculated with $\epsilon(\infty)$ are closer to the experimental
data.  Figure 4(b) represents identical calculations on zinc-blende
structure, with $m_e = 0.11$ and $m_h = 0.44$ \cite{Semi:92}, compared with
the experiments
\cite{Alivisatos:93,Hodes:87,Alivisatos:88,Park:90,Woggon:91,Nogami:91}.
In this case the exciton energies seem to be insensitive to
a reasonable choice  of the dielectric constant.

The bulk wurtzite CdSe semiconductor exhibits three distinct transitions,
labelled A, B, and C, arising from the splitting of the valence band
due to spin-orbit interaction.  If we assume that similar transitions
will exist in clusters too, then we obtain the transitions shown in
Fig. 4(c).
Since the bulk transition A is at 1.79 eV \cite {Semi:92}, and that
for a cluster of $R$ = 100 \AA~ is at 1.67 eV (using 233 plane waves),
a correction of +0.12 eV was applied.  The calculated transitions B
and C were also similarly corrected: B bulk transition = 1.81 eV
\cite{Semi:92}, cluster ($R$ = 100 \AA) = 1.74 eV, so the correction =
0.07 eV; C bulk transition = 2.22 eV \cite{Semi:92}, cluster ($R$ =
100 \AA) = 2.14 eV, so the correction = 0.08 eV.  From Fig. 4(c) we
observe that if only the lowest exciton energies are considered,
then the experimental data show significant scatter compared to
the calculated results.  However, when we consider all three transitions
(A, B, and C), then all the experimental data can be neatly accounted
for as belonging to one of these three transitions.   This finding
indicates the possibility that some of the experimental data
reported in the literature correspond to higher-energy transitions
(B and C) rather than to the lowest energy transition A.

The spectroscopy of nanoscale CdSe clusters have attracted considerable
attention from several research groups
\cite{H-Borr:87,Peyg:89,Brus:90,Eki:93,Baw:93,Alivisatos:93}.
Some of these groups have succeeded in
synthesizing CdSe cluster samples that have very narrow size
distributions \cite{Brus:90,Baw:93,Alivisatos:93}.
Furthermore, they have made
very careful measurements  of the exciton energies.
For these reasons, direct comparison between our
calculations and the data from these groups provides a benchmark test
of the reliability of our computational method in yielding accurate
exciton energies.  From the results presented above it is clear that
the agreement between our calculations and experiment is excellent.
The experimental data lie much closer to our calculated values than to
those of the EMM, without any exception.  The experimental measurements
on cluster samples with very narrow size distribution are in better
agreement with our calculations than the corresponding data
on samples with broad size distribution.  Finally, we are able to
identify that some measured exciton energies probably correspond to
higher energy B and C transitions than to the lowest energy A
transition.

\subsection{CdTe Clusters}

Since both Cadmium and Tellerium are large atoms, the CdTe clusters
and crystals exist
only in zinc-blende form.  Since both the cation
and anion are heavy atoms, we expect significant spin-orbit coupling
in CdTe.  Consequently, we have carried out
band structure calculations on zinc-blende CdTe clusters
using non-local EPM with the effects of  spin-orbit
coupling included in the Hamiltonian \cite{Chelikowsky:76}.   The
parameters used in these calculations are given  in Tables II and III.  The
spin-orbit coupling parameter $\mu$ was fit to the experimental band
gap  of 1.56 eV at 300 K \cite {Semi:92}.

Figure 5(a) presents the band structure for the zinc-blende CdTe
crystal, while Fig. 5(b) compares the calculated direct exciton
energies with the available experimental data
\cite{Rajh:93,Esch:90}.
In these calculations
we used $\epsilon$ = 10.2, $m_e$ = 0.09, and $m_h$ = 0.72
\cite{Semi:92}. The exciton energies of spherical clusters
of different radii are reported in Table VIII.
The computational procedure employed is identical to
that described above.  At present, experimental data on CdTe
clusters are few.  However, we hope that our calculations will
stimulate further experimental efforts on these clusters.

\subsection{AlP Clusters}

The binary semiconductor AlP is isoelectronic to Si$_2$ and they both
crystallize in an identical lattice structure.  Also, since Al and P
are the nearest neighbors of Si in the periodic table, it is reasonable
to expect that crystalline AlP will have electronic properties similar
to that of bulk Si.  Indeed that turns out to be true:  both Si and AlP
are indirect gap semiconductors with similar band structures.
Likewise, since both these semiconductors are made of light elements,
we expect spin-orbit coupling to be negligible in these crystals.  We
have reported on our investigation of the spectral shifts of silicon
clusters before.  Now we present corresponding calculations on AlP
clusters.

The previous calculated band structure of AlP was obtained using the
LCAO-MO method and the corresponding band structure is not available
from EPM calculations.  Consequently, we have fitted the non-local
parameters of AlP to reproduce the band structure calculated using the
LCAO-MO method.  Our initial guess for these parameters utilized the
parameters of bulk Si \cite{RK:91}.  In the end the optimal parameters
gave 3.63 eV for the direct and 2.41 eV for the indirect gaps.  These
calculated values are
in complete agreement with the corresponding experimental band gaps
at 300 K \cite{Madelung:91}.
Tables II and III give the parameters, while  Fig. 6
gives the bulk band structure and the calculated exciton
energies of AlP clusters.   These calculations utilized
$\epsilon$ = 9.8,  $m_e$ = 0.21, and $m_h$ = 0.94
\cite{Madelung:91,Sze:81}.  The exciton energies are reported in Table IX.
Nanoscale AlP clusters have not been synthesized in
the laboratory so far.
However, recent experimental efforts on Si clusters
point to the possibility of similar interest on AlP clusters.

\subsection{GaP and GaAs clusters}

Figures 7 and 8 give the band structures and the
exciton energies of GaP and GaAs spherical
clusters as a function of their radius. Tables II and III show the parameters
that were used  in these calculations \cite{Cohen:89}.
Figures 7 (a) and 8 (a) show the band structures of zinc-blende GaP and
GaAs crystals obtained using the non-local empirical pseudopotential
(solid line) and the local empirical pseudopotential (dashed line). Comparison
with the experimental data of GaP \cite{Ley:74,Eastman:74}
and GaAs \cite{Ley:74,Eastman:74,Aspnes:73,Chiang:80} crystals clearly shows
that the non-local approximation yields to  band structures that are better
in agreement with the experimental data than those one obtained with the local
approximation.
Since GaP is an indirect gap
semiconductor, we present spectral shifts of both the direct and
indirect transitions.
The results obtained from the use of both local
and non-local EPM are shown in these figures.
The calculated
band gaps of GaP, obtained using 283 plane waves and the local EPM
method, are 2.79 eV for the direct transition and 2.15 eV for the
indirect transition.  Since the corresponding experimental band gaps
are 2.78 eV for the direct transition and 2.27 eV for the indirect
transition \cite{Madelung:91}, we shifted the calculated band gaps by
-0.01 eV, and +0.12, respectively.  The non-local EPM gives 2.88
eV  and 2.17 eV, respectively, for the direct and the indirect band
gaps of GaP.  Consequently, the corresponding shifts are -0.10 eV and
+0.10 eV.

For GaAs, the experimental direct gap is 1.47 eV, which is an average
of the experimental band gaps at 0 and 300 K \cite{Madelung:91}.  The
calculated band gaps, obtained using 283 plane waves, are 1.50 eV with
the local calculation and 1.52 eV with the non-local calculation.
Consequently, the corrections are -0.03 eV and -0.05 eV,
respectively.  We used the bulk crystal parameters $\epsilon = 9.1$,
$m_e = 0.10$, $m_h = 0.86$ \cite{RK:91} for GaP clusters, and $\epsilon
= 9.1$, $m_e = 0.07$, $m_h = 0.86$ \cite{RK:91} for GaAs clusters.

The exciton energies thus calculated are presented in Tables X--XII
and Figs. 7 and 8.
 From these figures it is clear that over a large
range of cluster sizes the local EPM is able to reproduce spectral
shifts as accurately as non-local EPM for both GaP and GaAs.  However,
at small cluster sizes the non-local correction on the spectral shifts
is significant.  At large cluster sizes the absorption spectrum shifts
to higher energies with decreasing cluster size.  This blueshift is
expected due to confinement of the electron-hole pair in the cluster.
However, at small cluster sizes the absorption spectrum of GaAs
clusters shifts to lower energies with decreasing cluster size; a trend
opposite to that observed for large clusters.  In the case of GaP,
which is an indirect gap semiconductor, the lowest energy transition
exhibits the expected blueshift at both large and small cluster sizes.
But this
transition is not observable because it is forbidden.  The origin of
the absorption spectrum, corresponding to the observable direct
transition, shifts to lower energies with decreasing cluster size at
small cluster sizes.  Both local and non-local pseudopotentials exhibit
the same qualitative behavior.  The main difference is that the
non-local EPM predicts less redshift in small clusters than the local
EPM.

We can explain the calculated trends in the following way.  At large
cluster sizes the electron and hole are both confined in a spherical
well.  This quantum confinement increases the band gap with decreasing
cluster size and it is the dominant effect in this size regime.  In
these large clusters, the negatively charged electron and the
positively charged hole are spatially separated and hence the coulomb
attraction between them is negligible.  However, in small clusters the
coulomb attraction energy between the electron-hole pair cannot be
neglected.  While the band gap still increases with decreasing cluster
size, in small clusters this increase is sufficiently overcome by the
coulomb energy that the spectra shift to lower energies.  Consequently,
in this small cluster size regime the absorption spectra of clusters
may exhibit redshift instead of the blueshift.

At present reliable experimental data are not present for the spectral
shifts of these important III-V semiconductor clusters, partly because
of considerable experimental difficulties that arise in trying to
synthesize III-V semiconductor clusters with narrow size distribution.
However, based on our experience with CdS and CdSe clusters, and InP
clusters (see below), we expect our calculated spectral shifts of AlP,
GaP, and GaAs clusters to be good estimates of the expected spectral
shifts.

\subsection{InP clusters}

Recently, Nozik and co-workers have succeeded in synthesizing InP
clusters and for the first time showed an exciton transition in the
spectrum of a III-V quantum dot \cite{Nozik:94}.  The present work
compares their experimental results with  our theoretical predictions.  We
used the same procedure as before: First, we found the parameters of
the EPM Hamiltonian and then calculated the excitonic energies of InP
clusters.  The band structure calculations on zinc-blende InP crystal
were carried out using the non-local EPM that includes the effects of
spin-orbit coupling.  The parameters we employ are slightly different
from those used by Chelikowsky, but
otherwise the two calculations give nearly identical band structures.
In particular, the coupling parameter
$\mu$ = 0.00023 was chosen in order to reproduce the fundamental
splitting = 1.50 eV, as  obtained by Chelikowsky
\cite{Chelikowsky:76}.  Complete set of parameters used in our
calculations are tabulated in Tables II and III.

Figure 9(a) displays the band structure of bulk InP and Fig. 9(b)
presents the exciton energies of the clusters as a function of their
radius.  These calculations have employed the parameters $\epsilon$
= 9.61, $m_e$ = 0.077, and $m_h$ = 0.58 \cite{Madelung:91}.
We shifted all the conduction bands by -0.207 eV so that the band gap
of an $R$ = 100 \AA~ cluster is equal to the experimental bulk value of
1.35 eV at 300 K. The exciton energies are reported in Table XIII.

The type (a) sample in the experiments of Nozik and co-workers have a
radius of about 20 \AA~ and exhibits an exciton transition at about 1.7
eV \cite{Nozik:94}.
This paper quotes 0.35 eV blueshift
for the onset of absorption for the colloidal sample (a) whose mean
diamter is 26.1 \AA.  However, 0.35 eV is not the blueshift for a
particle with a diameter of 26.1 \AA~ because colloid (a) has a broad
size distribution with a standard deviation of 7.5 \AA~ (see Fig. 3 of
Ref. \cite{Nozik:94} ).
Thus the onset of absorption really corresponds to a size
of about 40 \AA~ (private communication).  We use this value for the
particle size while comparing experimental blueshift with our calculations.
The difference between our calculated exciton
energy and the experimental value is  approximately 0.17 eV.  The type
(c) sample has a radius of 13 \AA~ and exciton transition at 2.25 eV,
in perfect agreement with our calculation.  From this comparison it is
clear that the overall trend of the experimental spectral shifts is in
close agreement with our calculations.  Additional reliable data on
spectral shifts over a range of cluster sizes will certainly be useful
in establishing the accuracy of our calculations.

Our band structure calculations on all IIIA-VB
semiconductor clusters have shown that local empirical pseudopotentials
are reasonably accurate compared to the non-local empirical
pseudopotentials for the calculations of the spectral shifts of these
clusters.  The non-local corrections on the spectral shifts are most
important in small cluster sizes.  In addition, our calculations have
shown that, while quantum confinement energy is the dominant factor
affecting spectral shifts in large clusters, the Coulomb interaction
between the electron and hole has significant effect in small
clusters.  The attractive Coulomb interaction is sufficiently strong in
small clusters that it overcomes the confinement energy of the
electron-hole pair and gives rise to redshift, instead of the
blueshift, of the electronic absorption spectrum.

\section{Summary}
In summary, we calculated the spectral shifts of CdS, CdSe, CdTe, AlP, GaP,
GaAs, and InP semiconductor clusters using the most accurate
available empirical pseudopotentials.  These semiconductors
cover a wide range of bond polarities and  band structures.
These binary clusters
also represent a series in which one ion is held constant while the
other ion is varied along a column of the periodic table.
Furthermore, the pseudopotentials employed in these
calculations incorporate the effects of
non-locality and spin-orbit coupling whenever they are important.  For
many of these semiconductors, we had to first determine the
pseudopotential parameters based on the latest experimental and
theoretical data.  Consequently, we report complete set of
pseudopotential parameters employed in our calculations.  Furthermore,
we also give full details of our band structure calculations employing
pseudopotentials that incorporate both non-local and spin-orbit
effects.  At present these calculations represent the most
sophisticated calculations on the spectral shifts of semiconductor
clusters.

Previously we had shown that a simple local pseudopotential yields
spectral shifts of zinc-blende CdS clusters in excellent agreement with
experiment over a range of cluster sizes.  Now we extend these
calculations to include the effects of spin-orbit coupling in the
pseudopotential.  As before, we find that for zinc-blende CdS, wurtzite
CdS, wurtzite CdSe, and zinc-blende InP clusters the spectral shifts
calculated using our band structure model are in excellent agreement
with experiment.  The shapes and crystal structures of the unit cell
have significant effect on the exciton energies.   The small clusters
in particular are sensitive to whether their crystal structure is
zinc-blende or wurtzite.  In the case of small CdS clusters where the
experimental data on lattice structure are either ambiguous or
unavailable, our calculations are able to assign the structure
unambiguously.  While reliable
experimental data are not yet available at present for zinc-blende CdSe
clusters, we predict that the spectral shifts in these clusters will be
nearly the same as those of wurtzite clusters.

In the absence of experimental data on CdTe, AlP, GaP, and GaAs
clusters, our calculations provide reasonable estimates of the expected
spectral shifts and trends in these clusters as a function of cluster
size.  The very little difference between the local and the non-local
calculations confirms the validity of the local pseudopotential method
for the calculations of the exciton energies.  The main effect of the
non-local pseudopotentials seems to be to reduce the magnitude of the
redshift seen in the spectral shifts of some small III-V semiconductor
clusters.
Similarly, the spin-orbit interaction does not change the lowest exciton
energies.
The main effect of the
spin-orbit interaction is to split the valence band into sub-bands,
thus giving rise to new transitions in the spectra.  Some experimental
data seem to be in better agreement with these higher energy
transitions, thus suggesting the possibility that transitions
originating from spin-orbit split valence bands are being observed even
in these nanoscale clusters.  This also conceivably explains the
reasons for considerable scatter in the experimentally determined
exciton energies for a given cluster size.  This study, together with
our previous investigation, provides evidence that the empirical
pseudopotential method yields unique insights into the quantum
confinement effects and is a powerful tool for calculating the spectral
shifts of semiconductor clusters.

\section{Acknowledgments}

This research is supported by the New York University and the Donors of The
Petroleum Research Fund (ACS-PRF \# 26488-G), administered by the
American Chemical Society.  We are grateful to A. P. Alivisatos and
Moungi Bawendi for communicating their beautiful experimental data of
Fig. 4 in this paper.

\begin{table}
\caption{Reciprocal lattice vectors and form factors (in a.u.) for CdS
and CdSe crystals.}
\begin{center}
\begin{tabular}{cccccc}
${\bf G}$ & G$^{2}$ & ~~~~~~~~~~~~~~~~~~~~~~~~~CdS &  &
{}~~~~~~~~CdSe \\
&   & V$_{S}$ & V$_{A}$ & V$_{S}$ & V$_{A}$ \\ \hline
Zinc-blende \\
000 & 0 & 0.000 & 0.000 & 0.000 & 0.000 \\
111 & 3 & -0.120 & 0.115 & -0.115 & 0.095 \\
200 & 4 & 0.000 & 0.065 & 0.000 & 0.060 \\
220 & 8 & 0.015 & 0.000 & 0.005 & 0.000 \\
311 & 11 & 0.020 & 0.025 & 0.020 & 0.025 \\
222 & 12 & 0.000 & 0.025 & 0.000 & 0.025 \\
400 & 16 & 0.000 & 0.000 & 0.000 & 0.000 \\ \\
Wurtzite \\
000 & 0 & 0.000 & 0.000 & 0.000 & 0.000 \\
001 & $\frac{3}{4}$ & 0.000 & 0.000 & 0.000 & 0.000 \\
100 & 2$\frac{2}{3}$ & -0.130 & 0.000 & -0.125 & 0.000 \\
002 & 3 & -0.120 & 0.115 & -0.115 & 0.095 \\
101 & 3$\frac{3}{12}$ & -0.100 & 0.090 & -0.100 & 0.075 \\
102 & 5$\frac{2}{3}$ & -0.015 & 0.040 & -0.035 & 0.045 \\
003 & 6$\frac{3}{4}$ & 0.000 & 0.000 & 0.000 & 0.000 \\
210 & 8 & 0.015 & 0.000 & 0.005 & 0.000 \\
211 & 8$\frac{3}{4}$ & 0.000 & 0.000 & 0.000 & 0.000 \\
103 & 9$\frac{5}{12}$ & 0.020 & 0.025 & 0.015 & 0.025 \\
200 & 10$\frac{2}{3}$ & 0.020 & 0.000 & 0.020 & 0.000 \\
212 & 11 & 0.020 & 0.025 & 0.020 & 0.025 \\
201 & 11$\frac{5}{12}$ & 0.020 & 0.025 & 0.020 & 0.025 \\
004 & 12 & 0.000 & 0.025 & 0.000 & 0.025 \\
202 & 13$\frac{2}{3}$\ & 0.010 & 0.015 & 0.010 & 0.015 \\
104 & 14$\frac{2}{3}$\ & 0.000 & 0.010 & 0.000 & 0.010 \\
213 & 14$\frac{3}{4}$\ & 0.000 & 0.000 & 0.000 & 0.000
\end{tabular}
\end{center}
\end{table}

\begin{table}
\caption{The Pseudopotential parameters of the semiconductors.
The parameters $V_S$, $V_A$,
$\alpha_0$, and A$_2$ are in a.u.,  $a_0$ is in \AA,
and $\beta_0$ and $\mu$ are dimensionless.}
\begin{center}
\begin{tabular}{cccccccc}
Compound & $V_{S}(\sqrt{3})$ & $V_{S}(\sqrt{8})$ & $V_{S}(\sqrt{11})$ &
$V_{A}(\sqrt{3})$ & $V_{A}(\sqrt{4})$ & $V_{A}(\sqrt{11})$ & a$_{0}$ (\AA) \\
\hline
Local parameters \\
CdTe & -0.11000 & 0.00000 & 0.03100 & 0.03000 & 0.02500 & 0.01250 & 6.48 \\
AlP & -0.12250 & 0.01000 & 0.02000 & 0.05000 & 0.03000 & 0.00005 & 5.46 \\
GaP & -0.11500 & 0.01000 & 0.02900 & 0.05000 & 0.03500 & 0.01300 & 5.45 \\
GaAs & -0.10700 & 0.00700 & 0.03900 & 0.02800 & 0.01900 & 0.00100 & 5.65 \\
InP & -0.11750 & 0.00000 & 0.02650 & 0.04000 & 0.03000 & 0.01500 & 5.86 \\ \\
Non-local parameters \\
  &  & Cation &  &  & Anion &  & Spin Orbit \\
  & $\alpha_{0}$  & $\beta_{0}$ & A$_{2}$  & $\alpha_{0}$  &
$\beta_{0}$ & A$_{2}$  & $\mu$ \\
CdS  &   ...   &   ...   &   ...   &   ...   &    ...   &   ...   & 0.00008\\
CdSe &   ...   &   ...   &   ...   &   ...   &   ...    &   ...   & 0.00041\\
CdTe & 0.000 & 0.400 & 0.000 & 0.000 &  0.400 & 1.000 & 0.00061\\
AlP & 0.190 & 0.300 & 0.350 & 0.160 & 0.030 & 0.180 & ...\\
GaP & 0.000 & 0.300 & 0.200 & 0.160 & 0.050 & 0.225 & ...\\
GaAs & 0.000 & 0.000  & 0.063 & 0.000 & 0.000 & 0.313 & ...\\
InP & 0.000 & 0.250 & 0.275 & 0.150 & 0.050 & 0.175 & 0.00023
\end{tabular}
\end{center}
\end{table}

\begin{table}
\caption{The Pseudopotential parameters $R_0$ and $R_2$ in \AA.}
\begin{center}
\begin{tabular}{ccccccc}
Compound  &  & Cation & & & Anion &  \\
\hline
   & $R_{0}$  & & $R_{2}$  & $R_{0}$  & & $R_{2}$  \\
CdTe & 1.37 &  & 1.41 & 1.06  &   & 1.41 \\
AlP & 1.06 &  & 1.19  & 1.06 &  & 1.19 \\
GaP & 1.27 &   & 1.19  & 1.06 &  & 1.19  \\
GaAs & 1.27  &   & 1.44  & 1.06 &  & 1.44 \\
InP & 1.27 &  & 1.29 & 1.06  &  & 1.29
\end{tabular}
\end{center}
\end{table}

\begin{table}
\caption{Band gaps and exciton energies of zinc-blende CdS spherical clusters
obtained  using 283 plane waves.
$R$ is the radius of the cluster, $k$ is the angular momentum wave vector of
the lowest exciton state, $E_g$ is the band
gap, $V_c$ is the Coulomb energy, and $E_x$ is the
exciton energy. }
\begin{center}
\begin{tabular}{ccccc}
$R (\AA)$ & $k (2\pi/a_0) = a_0/\sqrt{12}R$ & $E_g$ (eV) &
$V_c$ (eV) & $E_x$ (eV) \\ \hline
5.0 &0.3359  &4.69  &-0.93  &3.77  \\
10.0 &0.1680  &3.69  &-0.46  &3.24  \\
15.0 &0.1120  &3.14  &-0.31  &2.85  \\
30.0 &0.0560  &2.65  &-0.15  &2.51  \\
50.0 &0.0336  &2.52  &-0.09  &2.44  \\
99.0 &0.0170  &2.46  &-0.04  &2.43
\end{tabular}
\end{center}
\end{table}

\begin{table}
\caption{Band gaps and exciton energies of wurtzite CdS spherical clusters
obtained  using 233 plane waves and neglecting the spin-orbit coupling.
$R$ is the radius of the cluster, $k$ is the angular momentum wave vector of
the lowest exciton state, $E_g$ is the band
gap, $V_c$ is the Coulomb energy, and $E_x$ is the
exciton energy. }
\begin{center}
\begin{tabular}{ccccc}
$R (\AA)$ & $k (\sqrt{2}\pi/a_0) = a_0/\sqrt{6}R$ & $E_g$ (eV) &
$V_c$ (eV) & $E_x$ (eV) \\ \hline
5.0 & 0.337700 &5.66  & -0.93 &4.66  \\
10.0 & 0.1689 &3.92   & -0.46 &3.39  \\
15.0 & 0.1126 &3.27   & -0.31 &2.89  \\
30.0 & 0.0563 &2.74   & -0.15 &2.52  \\
50.0 & 0.0338 &2.60  & -0.09 &2.45  \\
99.0 & 0.0171 &2.54  & -0.04 &2.43
\end{tabular}
\end{center}
\end{table}

\begin{table}
\caption{Band gaps and exciton energies of zinc-blende CdSe spherical clusters
obtained using 283 plane waves.
$R$ is the radius of the cluster, $k$ is the angular momentum wave vector of
the lowest exciton state, $E_g$ is the band
gap, $V_c$ is the Coulomb energy, and $E_x$ is the
exciton energy. }
\begin{center}
\begin{tabular}{ccccc}
$R (\AA)$ & $k (2\pi/a_0) = a_0/\sqrt{12}R$ & $E_g$ (eV) &
$V_c$ (eV) & $E_x$ (eV) \\ \hline
5.0 & 0.351000 &4.18  & -0.51 & 3.51 \\
10.0 & 0.1755 &3.18  & -0.25 &2.77  \\
15.0 & 0.1170 &2.64  & -0.17 &2.31  \\
30.0 & 0.0585 &2.14  & -0.08 &1.90  \\
50.0 & 0.0351 &2.00  & -0.05 &1.79  \\
99.0 & 0.0177 &1.94  & -0.02 &1.76
\end{tabular}
\end{center}
\end{table}

\begin{table}
\caption{Band gaps and exciton energies of wurtzite CdSe spherical clusters
obtained using  233 plane waves and neglecting the spin-orbit coupling.
$R$ is the radius of the cluster, $k$ is the angular momentum wave vector of
the lowest exciton state, $E_g$ is the band
gap, $V_c$ is the Coulomb energy, and $E_x$ is the
exciton energy. }
\begin{center}
\begin{tabular}{ccccc}
$R (\AA)$ & $k (\sqrt{2}\pi/a_0) = a_0/\sqrt{6}R$ & $E_g$ (eV) &
$V_c$ (eV) & $E_x$ (eV) \\ \hline
5.0 & 0.3510 &5.00  & -0.51 & 4.45 \\
10.0 & 0.1755  &3.28   & -0.25 &3.00  \\
15.0 & 0.1170 &2.60   & -0.17 &2.40  \\
30.0 & 0.0585 &2.04   & -0.08 &1.92  \\
50.0 & 0.0351 &1.88 & -0.05 &1.80  \\
99.0 & 0.0177 &1.81  & -0.02 & 1.76
\end{tabular}
\end{center}
\end{table}

\begin{table}
\caption{Band gaps and exciton energies of zinc-blende CdTe  spherical clusters
obtained using  137  plane waves and including the spin-orbit coupling.
$R$ is the radius of the cluster, $k$ is the angular momentum wave vector of
the lowest exciton state, $E_g$ is the band
gap, $V_c$ is the Coulomb energy, and $E_{x}$ is the
exciton energy.}
\begin{center}
\begin{tabular}{ccccc}
$R (\AA)$ & $k (2\pi/a_0) = a_0/\sqrt{12}R$ & $E_g$ (eV) &
$V_c$ (eV) & $E_x$ (eV) \\ \hline
5.0 &0.3741  &3.40  &-0.50  &2.89  \\
10.0 &0.1871 &3.00 &-0.25  &2.75   \\
15.0 &0.1247 &2.53 &-0.16  &2.36   \\
30.0 &0.0624 &1.94 &-0.08  &1.86   \\
50.0 &0.0374 &1.75 &-0.05  &1.69   \\
99.0 &0.0189 &1.65 &-0.02  &1.62
\end{tabular}
\end{center}
\end{table}

\begin{table}
\caption{Band gaps and exciton energies of zinc-blende AlP  spherical clusters
obtained using  283  plane waves.
$R$ is the radius of the cluster, $k$ is the angular momentum wave vector of
the lowest exciton state, $E_{xd}$ is the direct
exciton energy, and $E_{xi}$ is the indirect exciton energy. }
\begin{center}
\begin{tabular}{cccc}
$R (\AA)$ & $k (2\pi/a_0) = a_0/\sqrt{12}R$ & $E_{xd}$ (eV) &
$E_{xi}$ (eV) \\ \hline
5.0 &0.3154  &3.91  & 3.83 \\
10.0 &0.1577 &4.02  & 2.93  \\
15.0 &0.1051 &3.97  & 2.68  \\
30.0 &0.0526 &3.73  & 2.46  \\
50.0 &0.0315 &3.63  & 2.41  \\
99.0 &0.0159 &3.58  & 2.39
\end{tabular}
\end{center}
\end{table}

\begin{table}
\caption{Band gaps and exciton energies of zinc-blende GaP spherical clusters
obtained using 283  plane waves and the local pseudopotential.
$R$ is the radius of the cluster, $k$ is the angular momentum wave vector of
the lowest exciton state, $E_{xd}$ is the exciton energy  for the direct
transitions, and $E_{xi}$ is the exciton energy for the indirect transitions.}
\begin{center}
\begin{tabular}{cccc}
$R (\AA)$ & $k (2\pi/a_0) = a_0/\sqrt{12}R$ & $E_{xd}$ (eV) &
$E_{xi}$ (eV) \\ \hline
6.0 & 0.2624 & 3.09  & 3.15  \\
12.0 & 0.1312 & 3.43 & 2.58 \\
18.0 & 0.0875  & 3.25 & 2.43 \\
30.0 & 0.0525 & 2.99 & 2.31 \\
51.0 & 0.0309 & 2.84 & 2.26 \\
99.0 & 0.0159 & 2.79 & 2.25
\end{tabular}
\end{center}
\end{table}

\begin{table}
\caption{Band gaps and exciton energies of zinc-blende GaP spherical clusters
obtained using 283 plane waves and the non-local pseudopotential.
$R$ is the radius of the cluster, $k$ is the angular momentum wave vector of
the lowest exciton state, $E_{xd}$ is the exciton energy  for the direct
transitions, and $E_{xi}$ is the exciton energy for the indirect transitions.}
\begin{center}
\begin{tabular}{cccc}
$R (\AA)$ & $k (2\pi/a_0) = a_0/\sqrt{12}R$ & $E_{xd}$ (eV) &
$E_{xi}$ (eV) \\ \hline
6.0 & 0.2624 & 3.50  & 3.36  \\
12.0 & 0.1312 & 3.57 & 2.67 \\
18.0 & 0.0875  & 3.33 & 2.49 \\
30.0 & 0.0525 & 3.05 & 2.36 \\
52.0 & 0.0303 & 2.89 & 2.31 \\
98.0 & 0.0152 & 2.84 & 2.30
\end{tabular}
\end{center}
\end{table}

\begin{table}
\caption{Band gaps and exciton energies of zinc-blende GaAs spherical clusters
obtained using 283 plane waves.
$R$ is the radius of the cluster, $k$ is the angular momentum wave vector of
the lowest exciton state, $E_{xl}$ is the exciton energy  for the direct
transitions in a local calculation, and $E_{xnl}$ is the direct exciton energy
in a non-local calculation.}
\begin{center}
\begin{tabular}{cccc}
$R (\AA)$ & $k (2\pi/a_0) = a_0/\sqrt{12}R$ & $E_{xl}$ (eV) &
$E_{xnl}$ (eV) \\ \hline
6.0 & 0.2718 & 2.29  & 2.42  \\
12.0 & 0.1359 & 2.45 & 2.40 \\
18.0 & 0.0906  & 2.14 & 2.11  \\
30.0 & 0.0544 & 1.76 & 1.75 \\
51.0 & 0.0320 & 1.54 & 1.54 \\
99.0 & 0.0165 & 1.44 & 1.44
\end{tabular}
\end{center}
\end{table}

\begin{table}
\caption{Band gaps and exciton energies of zinc-blende InP  spherical clusters
obtained using  137  plane waves and including the spin-orbit coupling.
$R$ is the radius of the cluster, $k$ is the angular momentum wave vector of
the lowest exciton state, $E_g$ is the band
gap, $V_c$ is the Coulomb energy, and $E_x$ is the
exciton energy. }
\begin{center}
\begin{tabular}{ccccc}
$R (\AA)$ & $k (2\pi/a_0) = a_0/\sqrt{12}R$ & $E_g$ (eV) &
$V_c$ (eV) & $E_x$ (eV) \\ \hline
5.0 &0.3384  &3.12  &-0.53  &2.37  \\
10.0 &0.1692 &2.89 &-0.26  &2.42   \\
15.0 &0.1128 &2.50 &-0.17  &2.12   \\
30.0 &0.0564 &1.90 &-0.08  &1.60   \\
50.0 &0.0338 &1.67 &-0.05  &1.41   \\
99.0 &0.0171 &1.55 &-0.02  &1.32
\end{tabular}
\end{center}
\end{table}

\begin{figure}
\caption{
a) Band structure of
a wurtzite  CdS crystal implementing  the spin-orbit interactions
and 135 plane waves.
b) Allowed electronic
levels of wurtzite CdS spherical clusters ( radius = 15 \AA) obtained using
233
plane waves.
c) Band structure of a zinc-blende CdS crystal
obtained using 283 plane waves. Experimental data
are superimposed on the band structure for comparison [44].}
\end{figure}

\begin{figure} \caption{
a) Band structure of a wurtzite CdSe crystal implementing  the spin-orbit
interactions and 135 plane waves.
b) Band structure of a zinc-blende CdSe crystal obtained using
283 plane waves. Experimental data
are superimposed on the band structure for comparison [44].}
\end{figure}

\begin{figure}
\caption{ Direct exciton energy of  wurtzite (upper line), and
zinc-blende (lower line) CdS spherical
clusters compared with the experiments. Filled diamonds are used for
the zinc-blende data points while filled circles represent the wurtzite
clusters.
 233 plane waves were used
for the wurtzite  and 283 plane waves for the zinc-blende structures.}
\end{figure}

\begin{figure}
\caption{a) Direct exciton energies of wurtzite CdSe  clusters
obtained using two different dielectric constants: $\epsilon(0)$ (dashed line),
and
$\epsilon$($\infty$) (solid line). Comparison is made with the available
experimental data: stars [49]; plus [50]; squares [51]; up triangles
(light-hole data
from [52]); down triangles (split-off-hole data from
[52]); diamonds [53]; circles (SAXS data from [54]);
right triangles (TEM data from [54]).
b) Direct exciton energies of zinc-blende CdSe clusters
obtained using two different dielectric constant: $\epsilon(0)$ (dashed line),
and
$\epsilon$($\infty$) (solid line). The comparison is made with the available
experimental data: up triangles (SAXS data from [54]); down triangles
(TEM data from [54]); circles [55]; squares [56]; star [57]; plus [58];
diamonds [59].
c) Calculated A (lower line), B (intermediate line), C (upper line)
transition energies, as
discussed in the text. The experimental data are represented as in (a).
 }

\end{figure}

\begin{figure}
\caption{a) Band structure of zinc-blende CdTe crystals obtained using
137 plane waves and the spin-orbit interactions.
b) Direct exciton energies of zinc-blende CdTe clusters compared
with the experiments: circles [61]; diamonds [62].}
\end{figure}

\begin{figure}
\caption{a) Band structure of a zinc-blende AlP crystal obtained using
283 plane waves.
b) Direct exciton energies of zinc-blende AlP clusters obtained using
283 plane waves.
c) Indirect exciton energies of zinc-blende AlP clusters obtained using
283 plane waves.
} \end{figure}

\begin{figure}
\caption{a) Band structure of a zinc-blende GaP crystal obtained using 283
plane waves..
It is shown a
comparison between the non-local (solid line) and the local (dashed line)
empirical approximation. Experimental data are superimposed on the band
structure: diamonds [66]; plus symbols [67].
The non-local pseudopotential calculations
are seen to agree better with the experimental data than the local
pseudopotential calculations.
b) Direct exciton energies of zinc-blende GaP clusters.
The open circles are obtained using local pseudopotentials and filled
circles are obtained using non-local pseudopotentials.
c) Indirect exciton energies of zinc-blende GaP clusters.
The open circles are obtained using local pseudopotentials and filled circles
are obtained using non-local pseudopotentials.}
\end{figure}

\begin{figure}
\caption{a) Band Structure of a zinc-blende GaAs crystal. It is made a
comparison between the non-local (solid line), and the local (dashed line)
empirical approximation. Experimental data are superimposed on the
band structure: diamonds [66]; plus symbols [67]; $\times$ symbols [68];
squares [69].
The non-local pseudopotential calculations agree better with the experimental
data than the local calculations.
b) Direct exciton energies of GaAs spherical clusters.
The open circles are obtained using local pseudopotentials and filled circles
are obtained using non-local pseudopotentials.
The number of plane waves used
is 283.}
\end{figure}

\begin{figure}
\caption{a) Band structure of a zinc-blende InP crystal obtained using
137 plane waves and spin-orbit interactions.
b) Direct exciton energies of zinc-blende InP clusters compared with
experiments: circles [70].}
\end{figure}


\end{document}